\documentclass[aps,prl,reprint,superscriptaddress]{revtex4-2}
\usepackage{amsmath,amssymb}
\usepackage{graphicx,float}
\usepackage{times,txfonts}
\usepackage{color}
\usepackage{multirow}
\usepackage{bbold}
\usepackage{color}
\usepackage{soul}
\usepackage{subfigure}

\usepackage{hyperref}
\usepackage{times,txfonts}
\usepackage{float}
\usepackage{natbib}
\usepackage{nicefrac}
\usepackage{blindtext}
\usepackage[export]{adjustbox}
\usepackage{mathrsfs}
\usepackage{textcomp}
\usepackage{blkarray}
\usepackage{multirow}
\usepackage{ragged2e}

\begin{document}

\title{Efficient and high-fidelity entanglement in cavity QED without high cooperativity}



\author{S. Goswami}
\affiliation{Institute of Atomic and Molecular Sciences, Academia Sinica, Taipei City, Taiwan}

\author{C.-H. Chien}
\affiliation{Institute of Atomic and Molecular Sciences, Academia Sinica, Taipei City, Taiwan}

\author{N. Sinclair}
\affiliation{John A. Paulson School of Engineering and Applied Science, Harvard University, 29 Oxford St., Cambridge, MA 02138, USA}
\affiliation{Division of Physics, Mathematics and Astronomy, and Alliance for Quantum Technologies (AQT), California Institute of Technology, 1200 E. California Blvd., Pasadena, California 91125, USA}

\author{B. Grinkemeyer}
\affiliation{Department of Physics, Harvard University, Cambridge, MA 02138, USA}

\author{S. Bennetts}
\affiliation{Institute of Atomic and Molecular Sciences, Academia Sinica, Taipei City, Taiwan}

\author{Y.-C. Chen}
\affiliation{Institute of Atomic and Molecular Sciences, Academia Sinica, Taipei City, Taiwan}

\author{H. H. Jen}
\affiliation{Institute of Atomic and Molecular Sciences, Academia Sinica, Taipei City, Taiwan}
\affiliation{Physics Division, National Center for Theoretical Sciences, Taipei City, Taiwan}

\date{\today}

\begin{abstract}
The so-called state-carving protocol generates high-fidelity entangled states at an atom-cavity interface without requiring high cavity cooperativity.
However, this protocol is limited to 50\% efficiency, which restricts its applicability. 
We propose a simple modification to the state-carving protocol to achieve efficient entanglement generation, with unit probability in principle. 
Unlike previous two-photon schemes, ours employs only one photon which interacts with the atoms twice - avoiding separate photon detections which causes irrecoverable probability loss.
We present a detailed description and performance evaluation of our protocol under non-ideal conditions. 
High fidelity of 0.999 can be achieved with
cavity cooperativity of only 34.
Efficient state-carving paves the way for large-scale entanglement generation at cavity-interfaces for modular quantum computing, quantum repeaters and creating arbitrary shaped atomic graph states, essential for one-way quantum computing.

\end{abstract}

\maketitle

Entanglement, the quintessential feature of quantum mechanics, is a cornerstone of quantum technologies.
Entanglement at qubit-photonic interfaces is crucial for scaling quantum computers through modular quantum computing~\cite{monroe2014large, monroe2016quantum, li2024high, bluvstein2024logical, bravyi2022future,  aghaee2025scaling, gambetta2022quantum}.
They also play a central role in 
quantum repeaters \cite{briegel1998quantum, duan2001long, knaut2024entanglement} and other quantum communication paradigms such as enhancing quantum sensing capabilities in areas like optical clocks and long-baseline telescopes \cite{komar2014quantum, gottesman2012longer}, and interconnecting quantum computers (separated by lossy channels)\cite{wehner2018quantum, caleffi2024distributed, main2025distributed}. 
An important challenge in realizing these technologies is the ability to generate and grow entanglement efficiently, particularly at the interface between light and matter qubits.
	 
The state-carving (SC) technique~\cite{sorensen2003probabilistic}, recently demonstrated in atom-cavity interfaces \cite{djordevic2021entanglement, welte2017cavity, samutpraphoot2020strong}, offers a promising approach to high-fidelity light-matter entanglement generation in cavity quantum-electrodynamics (QED). 
SC operates with low cavity cooperativity, but it is inherently limited to an entanglement generation probability of 0.5.
SC requires the separate sequential detection of two ancillary photons to projectively create one particular atomic Bell state.
The low probability arises due to the separate detection of the two photons~\cite{sorensen2003probabilistic}.
	 
In this Letter, we introduce a high-fidelity and efficient SC scheme that overcomes this probabilistic barrier without the requirement of high cooperativity. 
Our improved scheme generates one of several Bell states through a properly constructed measurement in multiple Bell bases. 
This is achieved using two passes of a single ancilla photon, which always projects the atomic state onto a Bell state, and is measured once at the end of the protocol, avoiding separate detections.
	 
Moreover, our scheme requires only the addition of simple optical components like mirrors and beam splitters. 
The enhanced entanglement generation probability allows rapid advancement in applications - such as practical (low cooperativity) cavity meditated gates in quantum repeater nodes, between quantum computers or for scaling atomic one-way quantum computers~\cite{hein2004multi, raussendorf2001one, walther2005experimental}. 
Iterating the SC protocol produces large-scale, high fidelity atomic entangled states such as arbitrary shaped 2D graph states and cluster states for one-way quantum computation. 
While the original SC protocol presents a scaling barrier owing to its inefficiency \cite{chien2024generating},
our efficient SC protocol mitigates this scaling barrier.

There are several other approaches for efficient cavity-meditated atomic gates \cite{welte2018photon, chen2015carving, pellizzari1995decoherence, kastoryano2011dissipative, borregaard2015heralded, grinkemeyer2025error, sorensen2003measurement, duan2005robust, evans2018photon, majer2007coupling, feng2002quantum}. 
Many of these proposals require high cavity cooperativity ($C$), a major bottleneck in cavity QED.
Their fidelity errors vary as $\mathcal{O}(C^{-\frac{1}{2}})$ or $\mathcal{O}(C^{-1})$. Our efficient SC protocol, being a photon scattering protocol (like~\cite{welte2018photon}), has smaller $\mathcal{O}(C^{-2})$ fidelity errors and hence demands significantly less $C$.
See `{\it Results}' and Supplementary Materials (SM) Note 6~\cite{supp} for more details.

\begin{figure}[t]
    \centering
    \begin{subfigure}{}
      \centering     \includegraphics[width=\linewidth]{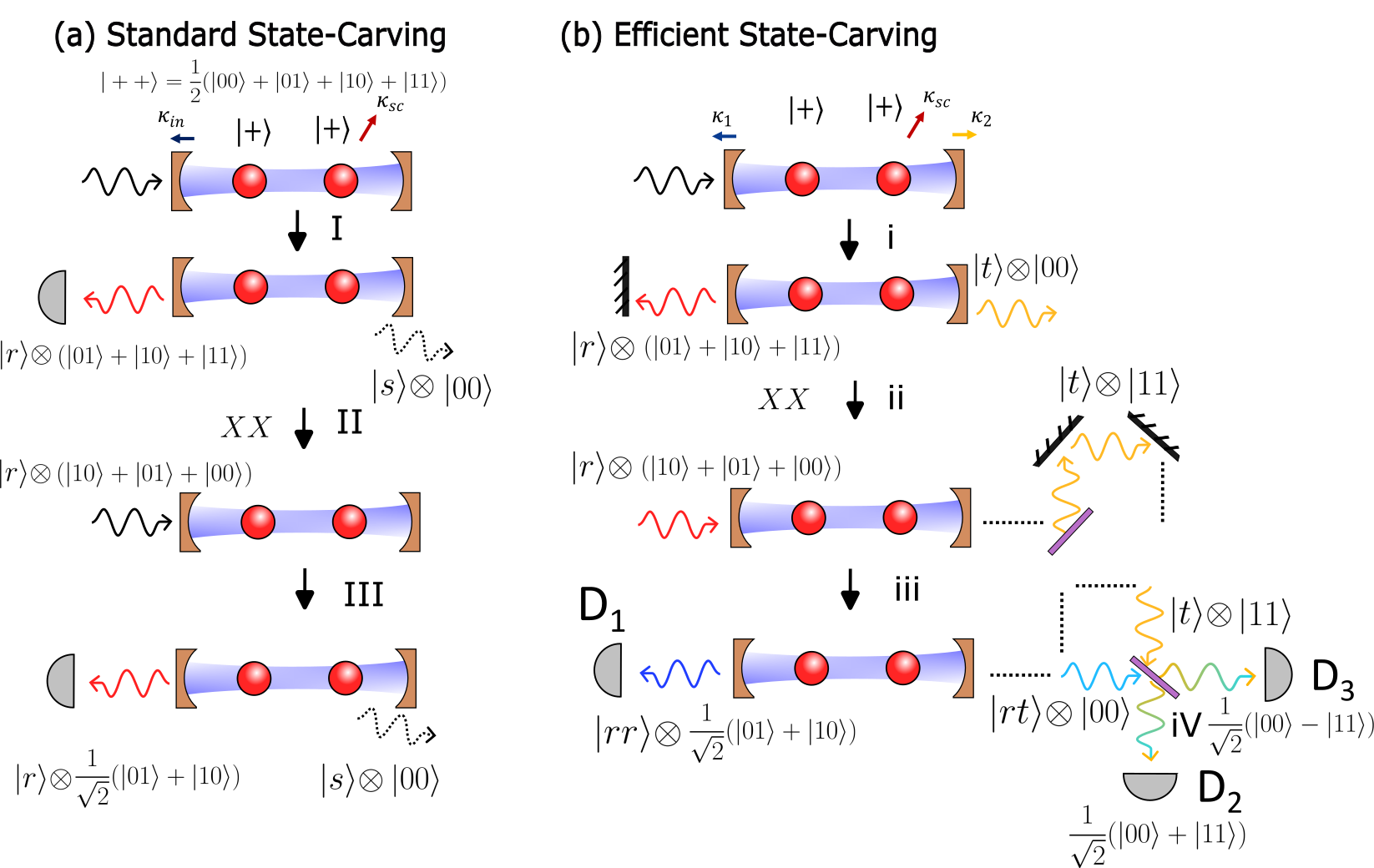}
    \end{subfigure}
    \begin{subfigure}{}
      \centering      \includegraphics[width=\linewidth]{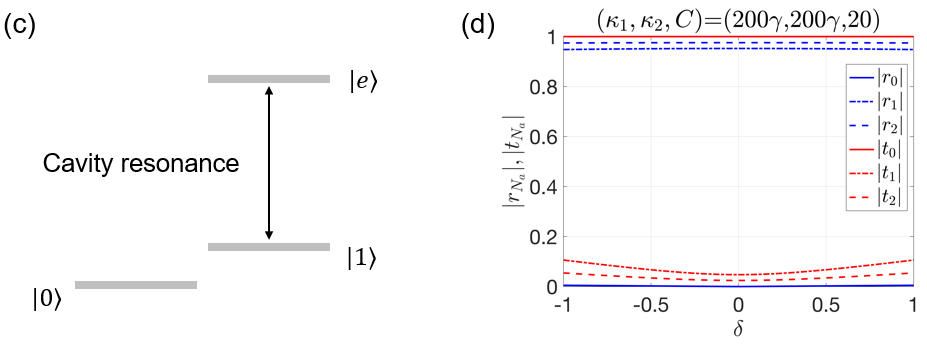}
    \end{subfigure}
    \caption{Schematic of the original state-carving (SC) protocol  (a) that ideally achieves 0.5 probability of atomic Bell-State generation and our efficient SC protocol (b) that achieves unit probability. Cavity loss channels ($\kappa_1$, $\kappa_2$, $\kappa_{sc}$ etc.) are shown at the top. (c) The atomic transition $|1\rangle$-$|e\rangle$ is coupled with the cavity while $|0\rangle$ remains uncoupled. (d) Reflection ($r_N$) and transmission ($t_N$) coefficients for the cavity coupled with $N$ atoms, with realistic cavity parameters $\kappa_1$ = $\kappa_2$ = 200$\gamma$~\cite{samutpraphoot2020strong, djordevic2021entanglement} and $C$ = 20.
    }
    \label{fig_1}
\end{figure}

\textit{Scheme}. The original state-carving (SC) protocol \cite{sorensen2003probabilistic} is shown in Fig. \ref{fig_1}(a). The evolution of the atom-photonic state through different stages of the protocol (I,II, III) is described in Eq. (\ref{Stand_SC}). An optical cavity couples to the $|1\rangle$-$|e\rangle$ atomic transition, leaving $|0\rangle$ uncoupled (Fig. \ref{fig_1}(c)). SC initially prepares two atoms in the separable state $|++\rangle = \frac{1}{2}(|00\rangle + |01\rangle + |10\rangle + |11\rangle)$. The atom-cavity interaction causes different reflectivity for the uncoupled state $|00\rangle$ (in principle 0) and coupled states $|01\rangle$, $|10\rangle$, and $|11\rangle$ (in principle 1). With a one-sided cavity containing equal input and scattering loss channels ($\kappa_{in} = \kappa_{sc} = \kappa/2$), an input photon scatters (to $|s\rangle$) for $|00\rangle$, and reflects (to $|r\rangle$) for coupled states \cite{djordevic2021entanglement}. Thus, input photon interaction with $|++\rangle$ yields an entangled atom-photon state (stage I). Detecting the reflected photon collapses the atomic state to $\frac{1}{\sqrt{3}}(|01\rangle + |10\rangle + |11\rangle)$ with 3/4 probability. To obtain the Bell state $\frac{1}{\sqrt{2}}(|01\rangle + |10\rangle)$, a second input photon and NOT gates on both qubits are used to eliminate $|11\rangle$ (by changing it to $|00\rangle$, at stage II). The second photon gets scattered if the atomic state is $|00\rangle$. If the second photon is reflected, detecting it yields the desired Bell state with 2/3 probability, giving an overall (3/4) $\times$ (2/3) = 1/2 probability (stage III). Note that the desired Bell state has a 0.5 overlap probability with the initial $|++\rangle$ state.

Our efficient scheme is shown in Fig. \ref{fig_1}(b) with the atom-photonic state evolution described in Eq. (\ref{Det_SC}). We avoid the probability loss of standard SC by using a single photon and a two-sided cavity - eliminating separate detection of $|00\rangle$ and $|11\rangle$ states while merging them to form entangled states $\frac{1}{\sqrt{2}}(|00\rangle \pm |11\rangle)$. Ideally, the cavity transmits an input photon when uncoupled (i.e., $|00\rangle$), and reflects it when coupled. Realistically, reflection and transmission coefficients differ slightly (Fig. \ref{fig_1}(d)), but the protocol retains high efficiency and fidelity. 

In the efficient protocol, the input photon - instead of being detected upon reflection from the cavity - is reflected back into the cavity using a mirror. 
The first pass of the input photon, with the two atoms initially in $|++\rangle$ states, enables the creation of an entangled atom-photonic states (Eq. (\ref{Det_SC}), stage i), where $|t\rangle$ and $|r\rangle$ represent the transmitted and reflected photonic states, respectively, after the first pass of the photon.
If the photon is reflected from the cavity, it is reflected back into the cavity using a mirror while applying NOT gates to the atoms (stage ii). 
Akin to the original SC scheme, the NOT gates flip $|11\rangle$ to $|00\rangle$ which is separated from the $|01\rangle$ and $|10\rangle$ states using the second pass of the reflected photon (stage iii). 
Labels $|rt\rangle$ and $|rr\rangle$ correspond to transmission and reflection on the second pass of the reflected photon (previously in $|r\rangle$ state). 
In stage iv, we interfere the two transmitted photons ($|t\rangle$ and $|rt\rangle$) on a 50/50 beam-splitter, to create Bell states $\frac{1}{\sqrt{2}}(|00\rangle \pm |11\rangle)$. Here, $ |t_{D_2}\rangle$ $\big(|t_{D_3}\rangle\big)  = \frac{1}{\sqrt{2}}(|t\rangle \pm |rt\rangle)$ with $|t_{D_2}\rangle$\big($|t_{D_3}\rangle\big)$ being transmission towards detectors $D_2\big(D_3\big)$, while $|rr\rangle$ state towards detector $D_1$ is renamed as $|rr_{D_1}\rangle$.
This is further described in SM Note 2~\cite{supp}.
The photon is detected with 0.5 probability in reflection and 0.25 each in the two transmission detectors. 
On measuring the photons, the atoms are projected into Bell states, ideally achieving probability 1 for Bell-state carving.
By using single-qubit gates conditioned on photon detection results one can create a particular Bell state (say, $\frac{1}{\sqrt{2}}(|01\rangle + |10\rangle)$), irrespective of which detector clicked.

In the efficient SC scheme (Fig. \ref{fig_1}(b)), the user chooses for the input photon to be reflected,  transmitted, or detected at different times when the photon travels along the same path.
To accomplish this in an experimental setup, multiple (three) optical switches may seem necessary. 
Although, loss in optical switches can lower the probability, high-extinction ($>$20 dB) and low-loss ($\sim$0.2 dB, including facets) switches with suitable bandwidth ($>$10 GHz) are feasible~\cite{psiquantum2025manufacturable,hu2021high,ying2021low}.

\begin{figure}[htbp]
   \centering
   \begin{subfigure}{}
     \centering      \includegraphics[width=0.9\linewidth]{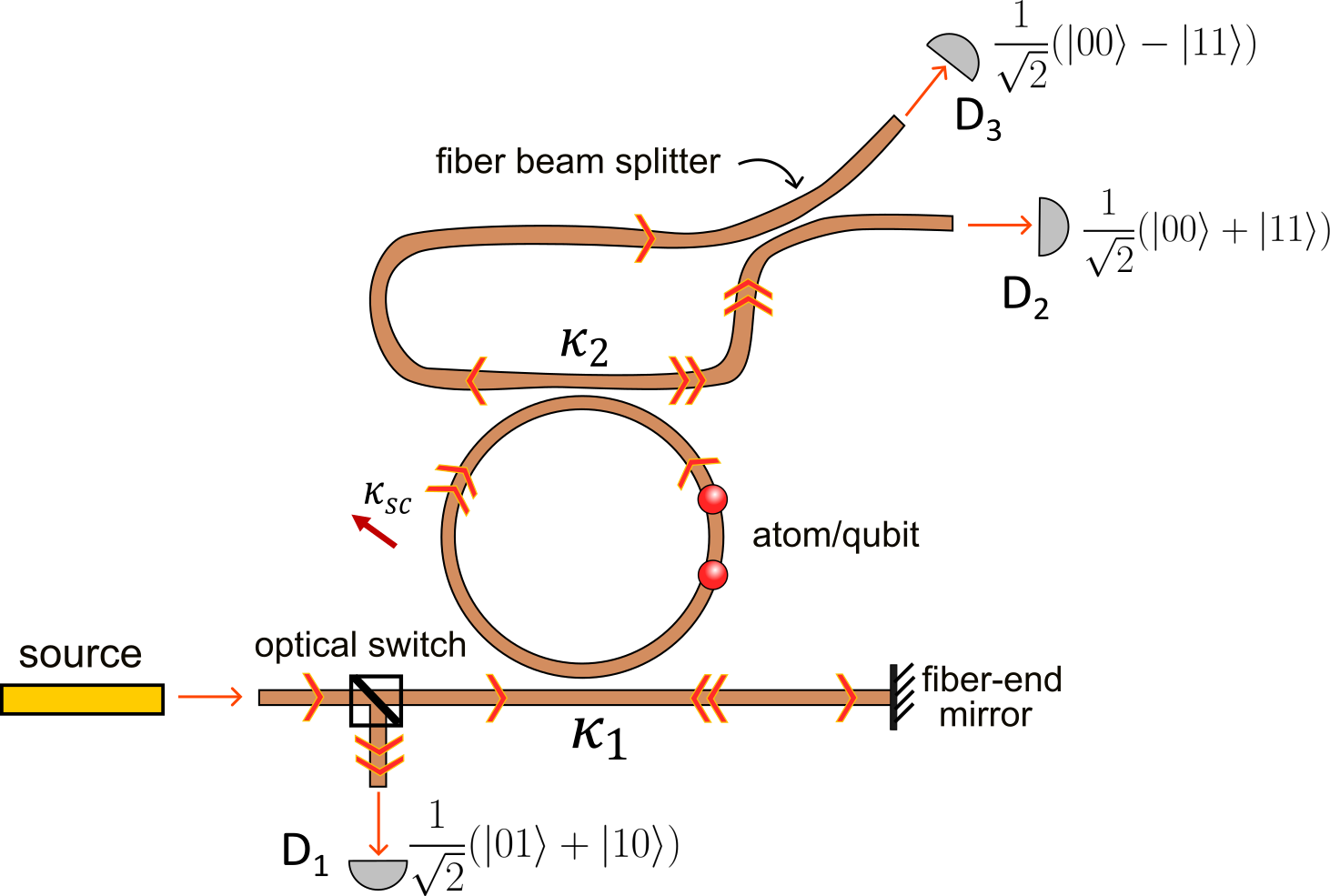}
    \end{subfigure}
    \caption{Implementation of the efficient state-carving protocol in a ring cavity - input photon (single arrow) is transmitted forward when `reflected' from the ring cavity and then gets reflected from the fiber-end mirror. The original input and the reflected photon (double arrow) couples to the cavity in different directions (counter-clockwise and clockwise respectively) and correspondingly transmits to the output fiber in different directions. While the two outputs are overlapped using a fiber beam-splitter, one optical switch (or isolator) is needed to detect the reflected input.
    }
    \label{fig_2}
\end{figure}

In a linear cavity, the scheme can be accomplished with one switch only if the photon is guided depending on its polarization, as described in SM Note 4~\cite{supp}. 
However, the scheme becomes much simpler in a toroidal or ring cavity \cite{strekalov2016nonlinear,cai2000observation, spillane2003ideality, mohageg2007high, zhuang2019coupling, shitikov2018billion}, as shown in Fig. \ref{fig_2}. 
Ring cavities does not need polarization guiding while still requiring only one switch (or an optical isolator \cite{jalas2013and}). 
In a ring cavity coupled to two optical fibers for input and output, `reflection' in the input fiber corresponds to forward transmission and light is coupled to different directions in output fiber depending on the direction of input photon. 

\begin{widetext}    
    \begin{eqnarray}\label{Stand_SC}
        \textbf{Standard SC : } |++\rangle \xrightarrow[]{\text{I}} \frac{1}{2}\bigg[|00\rangle|s\rangle + \bigg(|01\rangle + |10\rangle + |11\rangle\bigg)|r\rangle\bigg]
        \xrightarrow[]{\text{II}} 
        \frac{1}{\sqrt{3}}\bigg[|01\rangle + |10\rangle) + |00\rangle \bigg]
        \xrightarrow[]{\text{III}} 
        \frac{1}{\sqrt{3}}\bigg[\bigg(|01\rangle + |10\rangle\bigg)|r\rangle + |00\rangle |s\rangle\bigg] \hspace{0.1 cm}
    \end{eqnarray}
    \begin{eqnarray*} 
        \textbf{Efficient SC : } \hspace{0.1 cm} |++\rangle \xrightarrow[]{\text{\hspace{0.45cm} i \hspace{0.45cm}}} \frac{1}{2}\bigg[|00\rangle |t\rangle + \bigg(|01\rangle + |10\rangle + |11\rangle\bigg) |r\rangle\bigg]
        \xrightarrow[]{\text{{\hspace{0.45cm} ii \hspace{0.45cm}}}} \frac{1}{2}\bigg[|00\rangle |t\rangle + \bigg(|01\rangle + |10\rangle + |11\rangle\bigg) |r\rangle\bigg]
        \xrightarrow[]{\text{{\hspace{0.45cm} iii \hspace{0.45cm}}}}
    \end{eqnarray*}
    \begin{eqnarray}\label{Det_SC} 
        \frac{1}{2}\bigg[\bigg(|01\rangle + |10\rangle\bigg)|rr\rangle + |00\rangle|rt\rangle + |11\rangle|t\rangle\bigg]
        \xrightarrow[]{\text{{\hspace{0.45cm} iv \hspace{0.45cm}}}} 
        \frac{1}{\sqrt{2}}\bigg(\frac{|01\rangle + |10\rangle}{\sqrt{2}}\bigg)|rr_{D_1}\rangle + \frac{1}{2}\bigg(\frac{|00\rangle + |11\rangle}{\sqrt{2}}|t_{D_2}\rangle + \frac{|00\rangle - |11\rangle}{\sqrt{2}}|t_{D_3}\rangle\bigg)\hspace{0.3cm}
    \end{eqnarray}
\end{widetext}

\textit{Results}. With $N$ atoms coupled to the cavity, the ideal reflection ($r_N$) and transmission ($t_N$) coefficients are: $r_{0}=0$, $t_{0}=1$ and $r_{1}=r_{2}=1$, $t_{1} = t_{2}=0$. 
In all the discussions till now, we assumed this ideal scenario.
In a real experiment, deviations from the ideal scenario affect both state fidelity and entanglement generation probability. 
The non-ideal reflection and transmission coefficients for two-sided cavity are derived in the SM Note 1 \cite{supp}.

\begin{figure}[t]
    \centering      \includegraphics[width=\linewidth]{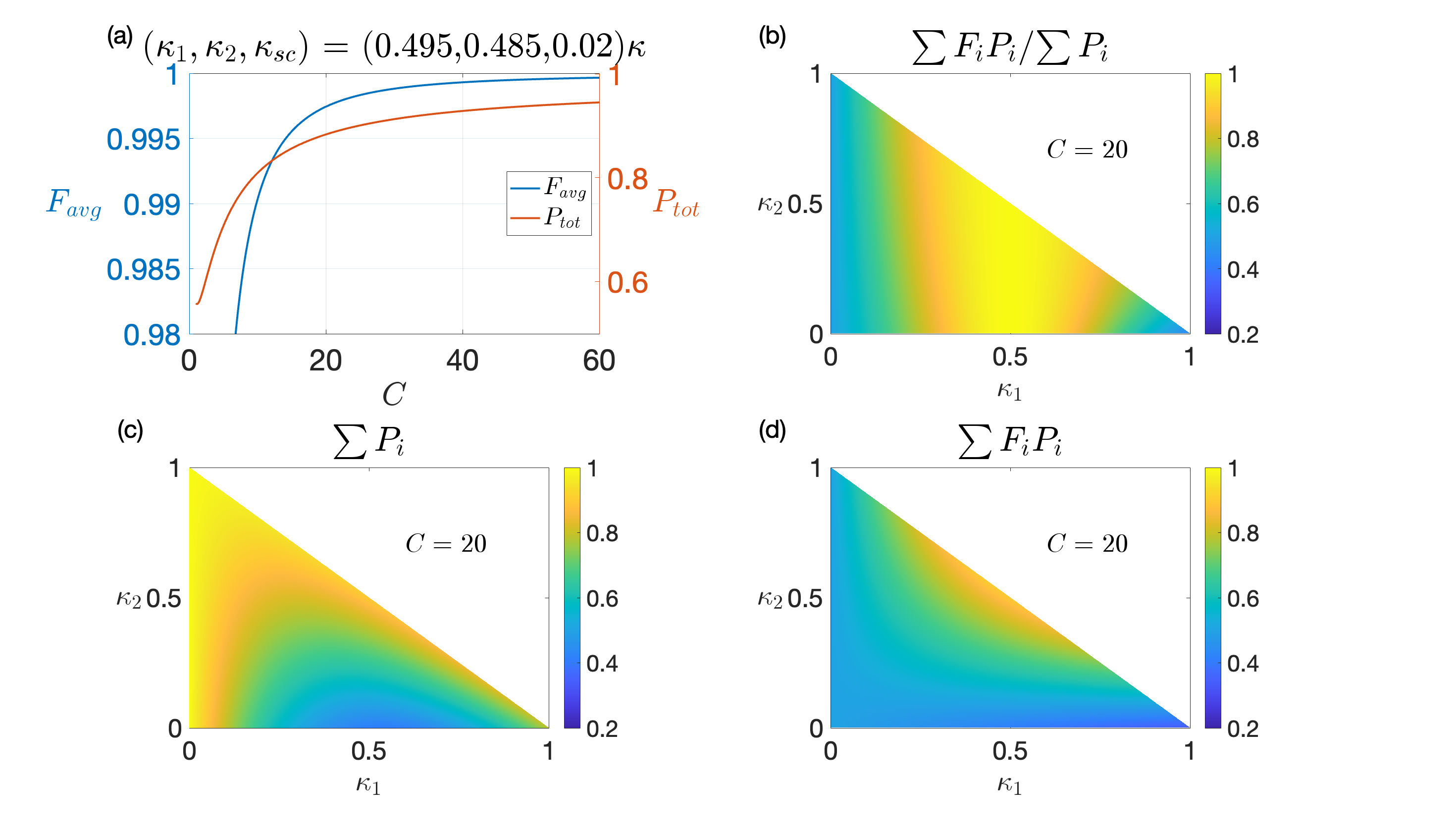}
    \caption{Fidelity and efficiency of efficient Bell state carving is shown with different cavity parameters. (a) High fidelity ($F =$ 0.999) can be achieved for moderate  cavity coperativity ($C \sim$ 34) while  efficiency (total probability) also nears 1 for high $C$ . (b)-(d) Normalized average fidelity in (b) depends almost only on $\kappa_1$ while total probability in (c) depends on both $\kappa_1$ and $\kappa_2$. In (d), un-normalized average fidelity highlights the ideal parameter regime for best performance in the determinstic SC proposal ($\kappa_1 \approx\kappa_2 \approx$ 0.5, $\kappa_{sc} \approx$ 0). We used $C$ = 20 for (b)-(d).
     }
    \label{fig_3}
\end{figure}

In a two sided cavity with couplings $\kappa_1$, $\kappa_2$ and scattering loss $\kappa_{sc}$ (see Fig. \ref{fig_1}(b)), the ideal scenario is realized when $\kappa_1 = \kappa_2 = \kappa/2$, $\kappa_{sc} = 0$ and very high cavity cooperativity $C = 4g^2/\kappa\gamma$ ($ C \xrightarrow{} \infty $) - with atom-cavity coupling $g$ and atomic linewidth $\gamma$. 
In general, neither of these two conditions are easily attainable in experiments. 
We show their impact on the performance of the protocol in Fig. \ref{fig_3}. 
In Fig. \ref{fig_3}(a), total probability ($P_{tot}$) and average fidelity ($F_{avg}$) are plotted with $C$. 
Total probability is given by $P_{tot} = \sum_i P_i$ and the average fidelity is given by $F_{avg}$ = $\frac{\sum^{3}_{i=1} F_i P_i}{\sum^{3}_{i=1} P_i}$, where $F_i$ and $P_i$ denote the fidelity and probabilities corresponding to the detections in three different detectors ($D_1, D_2, D_3$). $F_{avg}$ is normalized as opposed to Fig. \ref{fig_3}(d).
Please see SM Note 5~\cite{supp} for a description of the cases corresponding to the three different detectors. 

SC has the advantage of not requiring a high $C$ cavity, compared to  other cavity-mediated atomic gate proposals~\cite{welte2018photon, grinkemeyer2025error, duan2005robust, evans2018photon, majer2007coupling, feng2002quantum}. 
In standard SC~\cite{sorensen2003probabilistic, welte2017cavity, djordevic2021entanglement}, fidelity is ideally independent of $C$ (for $C \gg 1$, as $|01\rangle$ and $|10\rangle$ states have the same reflection coefficients irrespective of $C$) and
only the success probability depends on $C$, for which $C \sim $ 10 is enough (see Fig. 2 of \cite{sorensen2003probabilistic}). 
The same $C$ independence of fidelity is observed in our double reflected ($|rr\rangle$) case, as seen in Fig. S2(b) of SM~\cite{supp}.

In contrast, maximum achievable fidelity in many other cavity-mediated atomic gate proposals~\cite{welte2018photon, chen2015carving, pellizzari1995decoherence, kastoryano2011dissipative, borregaard2015heralded, grinkemeyer2025error, sorensen2003measurement, duan2005robust, evans2018photon, majer2007coupling, feng2002quantum} strongly depends on $C$. Gates based on photon exchange \cite{borregaard2015heralded, kastoryano2011dissipative, pellizzari1995decoherence} has fidelity varying as $F \sim1- \frac{1}{\sqrt{C}}$. The $\mathcal{O}(C^{-\frac{1}{2}})$ fidelity error can be improved by error detection through heralding to $\mathcal{O}(C^{-1})~$\cite{sorensen2003measurement, borregaard2015heralded, grinkemeyer2025error}. Further improvement in $C$ scaling of $F$ is not possible, but for particular cases fidelity can be improved by reducing the pre-factor~\cite{borregaard2015heralded}.

In our efficient SC protocol, the average fidelity ($F_{avg}$)  is analytically derived to be
\begin{equation}
    F_{avg}= \bigg(1-\frac{\epsilon_1^2}{2}\bigg)-\frac{17}{16}\cdot\frac{1}{C^2}
\end{equation}
where $\epsilon_1 = - r_0 = 1 - \frac{2\kappa_1}{\kappa}$, see SM Note 6.1~\cite{supp}. 
Our fidelity error depends on $C$ due to the transmitted cases, but it is a weaker $\mathcal{O}(C^{-2})$ dependence compared to the $\mathcal{O}(C^{-\frac{1}{2}})$ or $\mathcal{O}(C^{-1})$ dependencies of the photon exchange gates. 
A reason for this improvement is that our scheme is a heralded photon-scattering protocol which generally requires lower $C$ like in \cite{welte2018photon} (as explained in SM Note 6.3~\cite{supp}).
Fidelity of 0.999 is achieved for $C \sim$ 34 using our scheme (Fig. \ref{fig_3}(a)), while that requires $C \sim$ 1000 for typical $C^{-1}$ dependence.
Very recently, a different approach to efficient carving has been proposed to carve Dicke~\cite{dicke1954coherence} or GHZ~\cite{greenberger1989going} states in a few steps using Grover's algorithm~\cite{nagib2025efficient, nagib2025deterministic}, but with $\mathcal{O}(C^{-\frac{2}{3}})$ heralded fidelity errors.
Fidelity is also improved in a recent carving protocol~\cite{ramette2024counter}, while still being limited in success probability similar to the standard SC protocol.
However, our $F_{avg}$ is limited by $C$ only when  $\epsilon_1 \approx 0 $ (i.e., $\epsilon_1 \ll \frac{1}{C}$) or $\kappa_1 \approx \frac{\kappa}{2}$. 
$F_{avg}$ is upper bounded by $1 -\epsilon_1^2/2$. 
In Fig. \ref{fig_3}(a), $F_{avg}$ is bounded to maximum of 0.99995 for a symmetric cavity with $ \kappa_1 = 0.495\kappa$ and $\epsilon_1 = 0.01$.
There is always some scattering and absorption loss ($\kappa_{sc}$) in the cavity such that  $\kappa_1 + \kappa_2 + \kappa_{sc} = \kappa$. Finite scattering loss can be accommodated in our proposal without sacrificing fidelity by using an asymmetric cavity with $\kappa_1 \approx \frac{\kappa}{2}$ and consequently $\kappa_2 \approx \frac{\kappa}{2} - \kappa_{sc} < \kappa_1$. In Fig. \ref{fig_3}(a), $\kappa_{sc} = 0.02 \kappa$ is assumed (for 2$\%$ scattering loss) and hence $\kappa_2 = 0.485 \kappa$. 
Hence, very high fidelity can be reached in an asymmetric cavity even with finite $\kappa_{sc}$, as shown in detail in SM (see Note 6 and Fig. S5)~\cite{supp}.

Fidelity of all three individual final states (one reflected and two transmitted states) only depends on the ratio $\kappa_1/\kappa$ and C (not on $\kappa_2$ or $\kappa_{sc}$, as explained in SM Note 5.1~\cite{supp}). However, the probability of generating the entangled states depends on $\kappa_2$ too and therefore on $\kappa_{sc}$ ( = $\kappa – \kappa_1 – \kappa_2$) as well. In Fig. \ref{fig_3}(b), the normalized average fidelity ($F_{avg}$) is plotted with $\kappa_1$ and $\kappa_2$ ($C$ = 20). $F_{avg}$ depends mostly on $\kappa_1$ (slight dependence on $\kappa_2$ due to the presence of $P_i$ in $F_{avg}$). The maximum fidelity is reached around $\kappa_1 = \frac{\kappa}{2}$. The ideal case ($\kappa_1 = \kappa_2 = \frac{\kappa}{2}$) will evidently have perfect fidelity. However, $\kappa_2$ and $\kappa_{sc}$ doesn't really influence average fidelity much (and individual fidelity at all) and hence the condition for maximum fidelity will be very near $\kappa_1 = \frac{\kappa}{2}$, as seen in Fig. \ref{fig_3}(b).

We showed the variation of total probability ($P_{tot}$) with $\kappa_1$ and $\kappa_2$ in Fig. \ref{fig_3}(c), for C = 20. In SM Note 6.2~\cite{supp}, we showed $P_{tot} = 1 - \frac{\kappa_{sc}}{\kappa} - \frac{4}{4C}$. Hence, probability nears 1 for small $\kappa_{sc}$,  or $\kappa_1 + \kappa_2 \approx \kappa$. In some other regions, we also see high probability (e.g. when $\kappa_1$ $\sim$ 0). But, this is misleading. The corresponding generated states does not have high fidelity and may not even be entangled. In the extreme case of $\kappa_1$=0, the cavity is not even coupled to the incoming photon. Therefore, the input photon always gets reflected. Although the probability of detecting the reflected photon is 1, the corresponding atomic state is separable, having not interacted with the photon at all.

Hence, we introduced a new metric, un-normalized average fidelity or fidelity-weighted probability ($\sum_i F_i P_i$), to quantify the overall performance of our protocol. This metric can reach unity only if all generated states have perfect fidelity as the three probabilities sum below unity (considering the non-zero scattering probability corresponding to $\kappa_{sc}$). This new metric is plotted in Fig. \ref{fig_3}(d) with $\kappa_1$ and $\kappa_2$ for C = 20. We see that this metric reaches 1 only around the true ideal case of $\kappa_1 = \kappa_2 = \kappa/2$ for a large enough $C$ values.

\begin{figure}[t]
    \centering
    \begin{subfigure}
      \centering      \includegraphics[width=\linewidth]{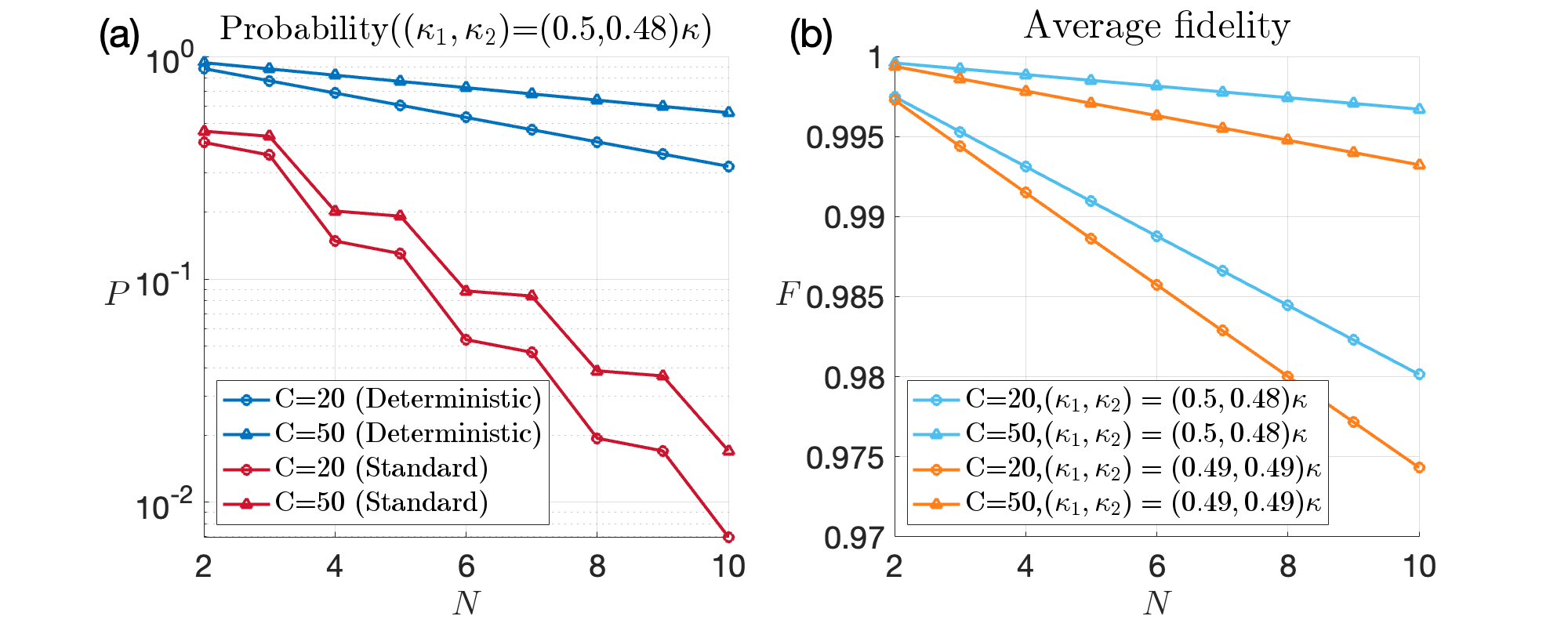}
    \end{subfigure}
    \caption{Probability of generation and fidelity of efficient state carving of large graph states (with $N$ nodes) is shown for cavity cooperativity, $C$ = 20 (circle), 50 (triangles). (a) For large N, generation probability of efficient SC protocol (blue) is orders of magnitude larger than that of standard SC protocol (red), which is shown for comparison.
    (b) Fidelity is principally dependent on C, with the asymmetric cavity (sky blue, $\kappa_1 \neq \kappa_2$) producing slightly better results than the symmetric cavity (orange).  
    }
    \label{fig_4}
\end{figure}

We described and analyzed a single efficient Bell-state carving protocol until now. 
Multiple state carvings can be combined by iterative addition of new qubits to create an arbitrarily-shaped large 2D graph state \cite{chien2024generating}. 
The probability of creating a large cluster state (with $N$ qubits) drops quickly though, if success probability for individual SC is limited to a maximum 0.5. 
Figure \ref{fig_4}(a) shows that the probability of generation is much larger for our efficient SC scheme compared to the original SC for large $N$, highlighting the principal advantage of this protocol. 
For standard SC, fidelity is perfect irrespective of $C$, if $\kappa_1$ = $\frac{\kappa}{2}$ \cite{chien2024generating}. 
Fig. \ref{fig_4}(b) depicts the average fidelity of the cluster states generated in efficient SC which is mainly limited by $C$.
The asymmetric ($\kappa_1$ = $\frac{\kappa}{2} > \kappa_2$) cavity parameters fares a bit better than symmetric ($\kappa_1$ = $\kappa_2$) ones as explained before (both with $\kappa_{sc} = 0.02$).
Arbitrary high fidelities can be achieved even in presence of finite cavity loss ($\kappa_{sc}$), because cases with lost photons are not detected and can be ignored. For the same reason, other losses like switch loss or detection loss only effect success probability, not fidelity.

Any arbitrary-shaped open graph state in 2D can be created by combining multiple two-qubit SCs. 
However, to create a general 2D cluster state, one may encounter closed loops or rings (e.g., in square cluster state or a box state \cite{chen2007experimental}). 
Closing a loop is a difficult task using SC. 
If controlled-Z gates are used to create cluster states, two-qubit operations would be sufficient to create a loop. 
However, because of the projective measurement nature of SC, one needs a three-qubit operation to close a loop \cite{chien2024generating}. 
If a certain state is discarded in a measurement, it cannot be recovered. 
Hence, SC on 3 qubits together is required to create arbitrary graph states containing closed loops.
It is possible to perform a efficient 3-qubit SC with unit probability by some modifications to the efficient 2-qubit SC setup. 
However, in such a two-qubit setup the probability of generation for the 3-qubit carving will be limited to 0.5. 
Both the efficient and 0.5 probability 3-qubit SC protocols are explained in detail in the SM Note 7~\cite{supp}.

We emphasize that we have not considered the effect of cavity coupling (or mode-matching) loss in our calculations. 
The coupling loss is an implementation issue that is present in nearly all protocols, including the original SC \cite{sorensen2003probabilistic}, although the coupling loss was cleverly bypassed in the experimental implementations of standard SC \cite{welte2017cavity,djordevic2021entanglement}. 
Near-perfect light-coupling efficiency has been achieved in whispering-gallery mode (WGM) micro-resonators \cite{strekalov2016nonlinear,cai2000observation, spillane2003ideality, mohageg2007high, zhuang2019coupling}, demonstrating couplings as high as $99.97\%$ \cite{spillane2003ideality}. 
The WGM ring resonators can also achieve reasonable co-operativity~\cite{will2021coupling, zhou2023coupling, pauls2020coupling} and very low scattering loss \cite{shitikov2018billion}. Importantly, the input coupling to WGM cavities (i.e., $\kappa_1$, which needs to be almost exactly $\frac{\kappa}{2}$) can be varied easily by adjusting fiber-cavity distance. Hence, WGM cavities could be overall an ideal system for implementation (also shown in Fig. \ref{fig_2}). 
Silicon vacancy centers in nanophotonic diamond cavities is another interesting implementation candidate which recently demonstrated high cooperativity and low mode-matching loss~\cite{bhaskar2020experimental, knaut2024entanglement}, which can be further improved using thin-film diamond~\cite{ding2025purcell}.

We presented a simple modification to the state-carving setup to achieve efficient entanglement generation ideally with unit probability, overcoming the limiting 50\% probability of the original SC protocol. 
The efficient protocol offers a pathway towards efficient, high-fidelity and practical (with low cooperativity cavity) entanglement generation at light-matter interfaces, which has significant implications for quantum technologies including quantum repeaters, modular quantum computing and scalable atomic one-way quantum computers. 

{\it Acknowledgments -} Authors thank Wei-Seng Hiew, Wun-Shan Zeng and Chun-Chi Wu, Faezeh Kimiaee Asadi, Mahsa karimi and Christoph Simon for fruitful discussions. HHJ and YCC acknowledge support from the National Science and Technology Council (NSTC), Taiwan, under the Grants Nos. 112-2112-M-001-079-MY3 and NSTC-112-2119-M-001-007, and from Academia Sinica under Grant AS-CDA-113-M04 and are also grateful for support from TG 1.2 of NCTS at NTU. 


%


\end{document}